\documentclass[aps, jcp, prb,  two column,showpacs,groupedaddress]{revtex4-1}
\usepackage{float}
\usepackage{graphicx} 
\usepackage{subfigure}
\usepackage{amsmath}
\usepackage{dcolumn}
\usepackage{multirow}
\usepackage{bm}     
\usepackage{gensymb}
\usepackage{amssymb,amssymb}   
\usepackage[left= 1 in, right= 1 in, top= 1 in , bottom= 1 in]{geometry}
\usepackage{hyperref}
\usepackage{xcolor} 
\hypersetup{ colorlinks, citecolor=blue ,linkcolor=black }

\usepackage{chemfig}

\usepackage{hyperref}

\hypersetup{linktocpage}

\begin{document}
	\vspace*{0.35in}

	\title{ Transport properties of Valine  in water at different temperatures}
	\author{Deepak Pandey}
	\author{ Narayan Prasad Adhikari }
	\email{npadhikari@tucdp.edu.np}
	\affiliation{Central Department of Physics, Tribhuvan University, Kirtipur, Kathmandu, Nepal}
	\begin{abstract}
			\addcontentsline{toc}{chapter}{Abstract} \noindent   Molecular Dynamics simulations of Valine in water  and their binary mixtures ($N_{Val}$ =0.003 \& $N_{water}$=0.997, $N$ representing the mole fraction) have been accomplished at temperatures 293.20 K, 303.20 K, 313.20 K, 323.20 K, and 333.20 K using the  OPLS/AA force field parameters. The work has been carried out by using GROMACS. The  OW-OW, H19-OW, N6-OW and  C1/C3-OW radial distribution functions (RDFs) have been estimated. Co-ordination numbers are also determined by the self-coded FORTRAN. The self-diffusion coefficients of Valine and water have been determined by means of mean-square displacement (MSD)  using Einstein's relation. The mutual diffusion coefficients of the binary mixtures have been determined using Darken's relation. The values of the diffusion coefficients have been found to agree with the experimental results within 8.54 \%. The temperature dependence of the diffusion coefficients have been analyzed and the analysis showed that they follow Arrhenius behavior. Energy estimated from Arrhenius plot agrees with experimental data within 13.04 \% for water and  5.34 \% for system.\\

	\end{abstract}
	
	\maketitle
	
	\section*{Introduction}
	\noindent  The amino acid contains both amine  (-NH$_2$)  and carboxylic acid (-COOH) as its  functional groups. Amino acid helps to make protein which is composed of C, O, N and H atoms.
	There exist twenty standard amino acids, out of them, nine are known as essential (indispensable) and rest are non-essential (dispensable). The essential amino acids  can't be produced by  our  human body\cite{dpk10}. 	Twenty different  L-$\alpha$-amino acids can build different sort of proteins, which include a linear polymer, where all the proteinogenic amino acids manifest mutual structural features with both $\alpha$-carbon to  the amino group, a carboxylic acid group. Amino acids can be classified also into polar (hydrophilic) and non-polar (hydrophobic)\cite{dpk10}. \\

\noindent Valine is symbolized by Val.  It is obtained by hydrolysis of protein. It is aliphatic R-group, long branched-chain amino acid (BCAA) and an essential (indispensable) amino acid which easily pairs with another BCCA (leucine, and isoleucine). Plants are the main source of essential amino acids. It is also known as {\it 2-amino-3-methyl butanoic acid}. It is a precursor in the penicillin bio-synthesis of proteins pathway \cite{dp}. It is expected to be found  the interior of  the protein.
Val is known as aliphatic, hydrophobic amino acid.
It's  molecular formula is C$_5$H$_{11}$NO$_2$.
Its molar mass is 117.148 g/mol,
 and the melting point  is 315 \degree C
\cite{lvaline}. \\

	\begin{figure}[h!]
		\centering
		\includegraphics[scale=0.38]{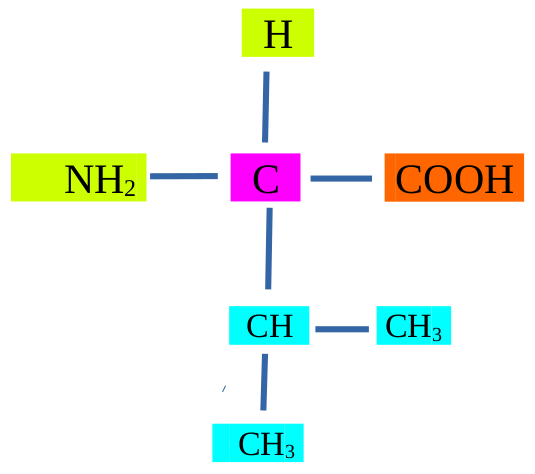}
		\caption{Molecular formula of Valine.}
		\label{fig31}
	\end{figure}
	
	\noindent FIG.\ref{fig31} shows a Val molecule. It contains $\alpha$-amino group (left side), an $\alpha$-carboxylic group (right side) and  methyl chain as (lower part),  Val in this structure  contains two non-hydrogen part to  C-beta carbon. It is bulkiness near to protein backbone, which streaks to adopt main-chain and an alpha-helical conformation; it lies within beta-sheets easily.  It helps to grow, regulate blood sugar, repairs muscles, cells, tissues, maintains the energy of our body as well as stir nervous system, and hels to proper mental functioning (mental vigor) etc. in our body\cite{bb}. It inhibits the transport of Tryptophan to the blood-brain barrier  which causes a  low level of mood, sleeps regulation of serotonin  \cite{dpk123}.
	Val helps to remove toxic (nitrogen) from  the liver which enables to transport nitrogen from one tissue of the body to other parts and maintain muscles as well as the regulation of the immune system.
	The deficiency of  Val can cause loss of weight, skin problems, hair loss, sleep disorder, mood swing, erectile disorders, arthritis, diabetes, maple syrup urine disorder, cardiovascular imbalance (high cholesterol level, high blood pressure, menopausal complaints) in human\cite{vro} and legs abnormalities in animals (mainly in chicken) \cite{dpk13}.   It  performs positive appetite suppressant, insomnia, and nervousness. Val plays  a vital role for muscles metabolism, where it assists right amount of Nitrogen in the body which provides stimuli. L-Val and D-Val isomers converted into glucose in  the liver is called  $Glucogenic$ $amino$ $acid$, which certainly balances emotional calm, and  muscles coordination. Val increases insulin secretion from the pancreas within 3--9 \%  and it boosts glycogen synthesis in  the muscle cell. Mostly wrestlers (bodybuilders) use Val with isoleucine and leucine
	to boost body growth and to regulate the energy \cite{x}.  L-Val uptake appears in the cotyledons when their water content has decreased to  65\% \cite{29} as well as the mechanism of amino acid transport in sea urchin embryos possess similarities to the well-described transport system present in mammalian\cite{1}. L-Val increases its urinary excretion in rats\cite{4,24}, the decreasing intestinal transport of L-Val with increasing age/weight in rats\cite{5}.  An experimental research has been carried out to calculate  the diffusion coefficient of amino acids  at the different concentration at 20 $^o$C in water \cite{12}. At each temperature diffusion coefficient  value decrease with increasing molar mass\cite{15}, diffusivity  gradually decreases with increasing interactive strength. \\

	\noindent Molecular dynamics simulations have been performed  to understand the physics behind reaction or extraction process of amino acids in our body, estimation of mass transfer rates is important: diffusion coefficients are one of the most fundamental quantities. Diffusion  can provide meaningful information about intermolecular interactions: the translation motion of the molecular diffusion reflects the friction arising from the intermolecular interaction between the surrounding \& diffusing molecules\cite{15}.  It is necessary and important to measure the diffusion coefficients of amino acids  from a scientific viewpoint  not only an engineering.
	 Despite the experimental estimation of the diffusion process of Valine~\cite{15}, to the best of our knowledge, there has been no molecular dynamics study on the diffusion of Valine in water at different temperatures.\\
	\section*{Diffusion}
	\noindent  The process of transferring  mass from  a system having the higher concentration to the region of a system having lower concentration as a result of random molecular motions is called diffusion~\cite{crank}. This phenomenon is chronicled by Thomas Graham ``Density especially does not play any role when different nature of gases brought into contact to arrange themselves like heaviest sink down and lighter float up,  but they  spontaneously diffuse, mutually and equally, through each other, and persist in the  pally state of the mixture for any length of time''.
		The sort of diffusion where  no chemical concentration gradient exists in a homogeneous system is called Self-   Diffusion. It often occurs when the mixing of identical molecules to the solution overall. The corresponding coefficient is called self-diffusion coefficient. 	Being the different characters in concentration  for  the same sort of molecules invites to  an equilibrium state, is  begotten by random motion. 	Self-diffusion is characterized due to noncooperative random-walk motion of the individual molecules \cite{abc}. This diffusion is studied by  Mean square displacement (MSD)  known as {\it Einstein relation}\cite{sunil}.  
	In case of three dimensional system,  the mean square displacement (MSD) and self-diffusion coefficient can be related by;
	 \begin{equation}
	 \label{Einstein}
	 D = \lim_{t\rightarrow\infty} \frac{\left\langle {[r(t)-r(0)]^2}\right\rangle}{6t}. 
	 \end{equation}
	 In equation \ref{Einstein} the bracket term $[r(t)-r(0)]$ represents change in position of diffusing particles at time $t$ from $t$ = $0$ and  nominator of this equation is ensemble average such that 
	 $[r(t)-r(0)]^2$ related with MSD of diffusing particles. When we draw the MSD vs time then we get a straight line then we fit best-fitted line, and slope of this best fitted line gives the value which is equal to the sixth times of diffusion coefficient of that particle at that temperature.\\
	 
\noindent	The sort of diffusion which occurs in  the binary mixture (two distinct molecules) i.e., the heterogeneous system is known as Binary Diffusion.
	The corresponding coefficient is called Binary Diffusion coefficient. In  various senses, we can call it mutual, chemical, inter transport --Diffusion. The expression for Binary Diffusion  coefficient  is given by famous Darken's relation \cite{aaa},
	\begin{equation}
	\label{darken}
	D_{12} = D_1N_2 + D_2N_1
	\end{equation}
		where $D_{12}$ is the binary diffusion coefficient. 
	$D_1$ \& $D_2$ are self-diffusion coefficients of the respective molecules.  
	$N_1$ \& $N_2$ are the  mole fraction of respective molecules.
	If the mixture is about to infinitesimal then the binary  diffusion coefficient is almost equal to the self-diffusion coefficient of one of the constitutive particles (components).  The theoretical analysis of binary diffusion is more complex than self-diffusion  due to being characterized via  cooperative  motion \cite{sunil}.

	\section*{Computational Details}

	\noindent
	The classical molecular dynamics (MD) is  one of the best computational simulation technique for equilibrium and transport property of   many ($N$) body problems.  If positions and velocities   of the state of molecules at any instead of time are well known  then it seems more predictable The classical method  that refers to the nuclear motion of the integral particles obeys the laws of classical mechanics- this is the brilliant estimation for an extensive sector of materials\cite{smith}.
	Using MD, Lennard-Jones potential can explain  van der Waals interactions which are based on Fritz London theory \cite{sunil}.
	MD is more similar to the real experiments: what we perform in real experiments we do same.\\
	
	\noindent   In MD simulation we can measure an observable quantity--which must be expressed as a function of position and momenta of the particles in the system.  The temperature in a classical $N$-body system, using equipartition of energy overall degrees of freedom that enter quadratically in the Hamiltonian of the system, average K.E. per degree of freedom,
	\begin{equation}
	<\frac{1}{2}m  v^2_\alpha  > = \frac{1}{2} k_B T.
	\end{equation}
		\noindent	   We solve Newton's equation for individual interacting ( $i^{th}$) particles  i.e.,
	\begin{equation} 
	\label{1.111}
	m_i \dot {\boldsymbol v}= -\nabla_iU({\bf r}_i) = {\bf F}_i, \hspace{0.7cm} \dot{\boldsymbol v} =\frac{\partial^2 {\bf r}_i}{\partial t^2}	
	\end{equation}

	\noindent here, $m_i$ is the mass, ${\bf r}_i$ is the position of the  $i^{th}$ particles, U({\bf r}$_i$) is the average potential experienced by the  $i^{th}$ particles and ${\bf F}_i$ is the mean force on the particle. 	\\
	
	\noindent We start  molecular dynamics process by 	modeling system which consists of specified molecules as atomic masses, charges, van der Waals radii, force field and more.  This model contains the topology-consists of atoms, which are connected to each other \& resembles with the  real system.  Atoms experience a force from its pairwise additive interaction with other systems. The total potential energy is linear contributions of bonded and non-bonded interactions~\cite{allen};
	 \begin{equation}
	  \small  {
	 \label{potentialenergy}
	 U_{total} =  \underbrace{( U_{bond} + U_{angle} + U_{dihed} )} +      \underbrace{ (U_{vdW} + U_{Coulomb})}}
	 \end{equation}
	\hspace{1.1cm}  \small{	 bonded potentials \hspace{.3cm}  non-bonded interaction}
	 	
	 \noindent Then, we specify the initial positions and velocities to all molecules in our input file ($val.pdb$) which enables to create a trajectory of the particle in the phase space. Velocities are  created randomly and rescaled to match  the constrained temperature. Different interactions are responsible for force experience by individual atoms. Force in each particle is calculated in each step and trajectory is created in accordance. Using Maxwell-Boltzmann distribution,    we can find the position and velocity of each particle and integrate Newton's equation of motion with leapfrog algorithm.
	\subsection*{Simulation Set Up} 
	
	\noindent Simulation was carried out for Val-water system at a different temperatures. Val (C$_5$H$_{11}$N$_2$)  is electrically neutral, it contains an amine group (NH$_2$) and a carboxyl group (COOH)  covalently bonded with C4. The 4 C -atoms are also covalently bonded as C-C-C-C. Because of the difference in electro-negativity, different atoms in this molecule have different partial charges. The structure of Val molecule depends strongly in the $pH$ of the medium and takes neutral form. 
	Bonds, bond angles, and  proper  dihedral characterizes its (Val's) topology and  electrostatic properties are led by the partial charges and van	der Waals interactions are controlled by defining LJ parameters. All the bonded and	non-bonded interactions  are taken into account in our model to	make the Val as real as possible. While modeling this, we have taken all atoms model so all atoms are characterized individually. OPLS/AA {\it oplsaa.ff} force field is used to optimize  all interactions in  the GROMACS package. Necessary information about the bond structures with respective {\it opls-code} and charges are saved  in the file  {\it aminoacids.rtp}   as well as  atoms types and their respective masses are elucidated in $atomtypes.atp$. We use $nrexcl$ $3$ to exclude non-bonded interaction in nearest three bonded neighbor.\\

	\noindent We have used the SPC/E water model. It accounts for the best values of the bulk water dynamics structure. It has the high value of dipole moment around 2.35 D and we know charge models are constant this correctly adds 1.25 kcal/mol to the total energy. This model results in better density and diffusion constant than SPC model. It gives accurate tetrahedral angle 109.49$\degree$ where other models give 104.52$\degree$. This model consists of 3 point charges on each atomic site, one  oxygen atom  pair with two  hydrogen atoms (HW1--OW--HW2). Although the water molecule is electrically neutral,  each atom carries partial charges where each hydrogen atom carries a partial charge of  +0.4238e, and oxygen atom carries a partial charge of -0.8476e. Here, e = 1.6022$\times$10 $^{-19}$ C is the elementary charge\cite{bhakta}. In GROMACS, the file \textit{spce.itp} contains all these parameters. The force field parameters for  the rigid SPC/E model are presented below:
	
	\begin{table} [H]
		\centering
		\label{forcefieldforH2O}
		\caption{Force field parameters for SPC/E water model.}
		\begin{tabular}{|c|c|} \hline
			Parameters &Values\\
			\hline
		$K_{OH}$ &  3.45  $\times$  $10^5$  kJ mol$^{-1}$nm$^{-2}$  \\
		$b_{OH} $ &  0.1 nm \\
		$K_{HOH}$ & 3.83   $\times$  $10^2$   kJ mol$^{-1}$rad$^{-2}$\\
		$\Theta_o $ & $ 109.47 \degree $\\
			\hline
		\end{tabular}
	\end{table}

	\noindent The information about the non-bonded parameters are explained in file \textit{ffnonbonded.itp} follows;\\
	
	\resizebox{0.45\textwidth}{!}{\begin{tabular}{ccccccc}
			[\textbf{atomtypes}]&& &&&&\\
			
			name &  at.num &   mass   &     charge &  ptype &        sigma ($\sigma$)  &      epsilon ($\epsilon$)\\
		 C7  &   6   &   12.01100 &   0.520   &   A & 3.75000e-01 & 4.39320e-01\\
		 O9    & 8    &  15.99940   & -0.440  &  A   & 2.96000e-01  & 8.78640e-01\\
		 O8   &  8     &   15.99940 &   -0.530   &   A &   3.00000e-01 & 7.11280e-01\\
		 H19   & 1     &  1.00800  &   0.450    &   A  &  0.00000e+00 & 0.00000e+00\\
		 N6   &  7      & 14.00670   & -0.900    & A  & 3.30000e-01 & 7.11280e-01\\
		 H18  &  1     &   1.00800   &  0.360   &  A  & 0.00000e+00 & 0.00000e+00\\
		 H17  &  1  &	 1.00800 &    0.360   &    A & 0.00000e+00 & 0.00000e+00\\
		 C4   &  6   &   12.01100  &  0.120  & A  & 3.50000e-01 & 2.76144e-01\\
		 H5   &  1  &	1.00800   &  0.060   &   A   & 2.50000e-01& 6.27600e-02\\
		 C2  &	6 &       12.01100  &  -0.060  &    A &  3.50000e-01 & 2.76144e-01\\
		 H13  &	1  &     1.00800   & 0.060    &   A   & 2.50000e-01  & 1.25520e-01\\
		 C3   &	6   &    12.01100   & -0.180   &  A   & 3.50000e-01  & 2.76144e-01\\
		 H10  &	1    &   1.00800   & 0.060   &   A   & 2.50000e-01   & 1.25520e-01\\
		 H11  &	1  &  1.00800   &  0.060   &   A  & 2.50000e-01 & 1.25520e-01\\
		 H12  &	1     &  1.00800  & 0.060  &   A  & 2.50000e-01 & 1.25520e-01 \\
		 C1  &	6    &   12.01100   & -0.180  &   A & 3.50000e-01 & 2.76144e-01\\
		 H14  &	1    &   1.00800    &  0.060   &   A  & 2.50000e-01  & 1.25520e-01\\
		 H15  &	1    &   1.00800    &  0.060    &   A  & 2.50000e-01  & 1.25520e-01\\
		 H16  &	1    &   1.00800    &   0.060  &  A   & 2.50000e-01  & 1.25520e-01\\
		 OW &   8 &   15.99940  &   -0.820 &   A  &   3.16557e-01 &    6.50194e-01\\   
		 HW1  & 1 & 1.00800 & 0.410 &  A &  0.00000e+00 &  0.00000e+00\\
		 HW2  & 1 & 1.00800 & 0.410 &  A &  0.00000e+00 &  0.00000e+00\\
		\end{tabular}}\\
		
\begin{figure}
			\centering
			\includegraphics[scale=0.35]{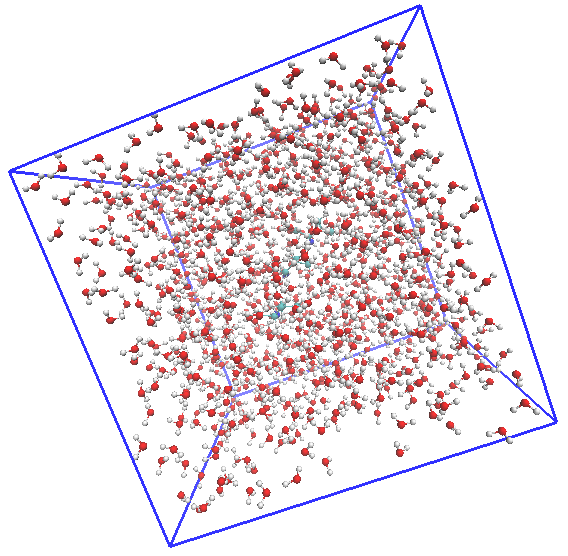}
			\caption{\small System under study after energy minimization.}
			\label{enm2}
		\end{figure}

		\noindent $1^{st}$ column represents the name of  corresponding atoms, as well as  OW stands for oxygen atom of water, HW1 and  HW2 stand for  two hydrogen atoms of water molecules. $2^{nd}$ column  notifies the atomic number of respective atoms of our system, $3^{rd}$ stands for their masses in the atomic unit. $4^{th}$ column specializes partial charges of atoms of the molecule. $5^{th}$ column  itemizes particle type and A stands for atom. $6^{th}$ and $7^{th}$ columns give the  value of $\sigma$ and $\epsilon$  of the corresponding atoms. GROMACS does not understand  $\sigma$ and $\epsilon$  such that we convert them into the form of $C_{ij}^6$ and $C_{ij}^{12}$ using;
			\begin{equation} 
			C^{(12)}_{ij} = 4 \epsilon_{ij} \sigma^{12}_{ij} \hspace{0.5cm} \&  \hspace{0.5cm}  C^{(6)} _{ij} = 4 \epsilon_{ij} \sigma^{6}_{ij}
			\label{1a} 
			\end{equation}
			\noindent The pairs which are involved in the LJ interaction and  modified parameters $\sigma_{ij}$ and $\epsilon_{ij}$  can be calculated by  Lorentz--Berthelot rules\cite{usermanual};
			\begin{eqnarray}
			\label{2.23234}
			\sigma_{ij} =  ( \sigma_{ii} \sigma_{jj})^{\frac{1}{2}} 
			\hspace{.5cm} 
			\& \hspace{.5cm}	\epsilon_{ij} = ( \epsilon_{ii} \epsilon_{jj})^{\frac{1}{2}} 
			\end{eqnarray}
		\noindent Before solvation, only three molecules of Val are presented in the cubic box of dimension ~3~nm~$\times$3~nm~$\times$3~nm.
	After completion of solvation of Val in water by \textbf{genbox} command. Now our system is ready for simulation.
	Still, the configuration of molecules specified in  \textit{afgen.gro} (structure file after solvation) is very far from being  an equilibrium: it might contain bad contacts, atoms nearer than van der Waal radius.  Forces may be too large and  this tends to fail the  MD simulation. To succeed  in this failure,  the process of energy minimization is required which helps to bring  our system in the equilibrium state. 
	In order to remove ``clashes" i.e., close over ions (LJ core) we perform energy  minimization, which  reduces the thermal noise  in the structures and potential energies so they can be compared better.  
	 The steepest-descent algorithm is used as the integrator for em, without fixing any constraints because it  moves in the direction parallel to the force (i.e., negative gradient of potential energy function) to reach the minimum; without  considering any of the history built in the past.  Here, {\it nsteps} is the maximum number of (minimization) steps to perform for this process. {\it emtol } = 50 represents the force tolerance in kJ mol$^{-1}$ nm$^{-1}$ means it 	helps to stop minimization when the maximum force  $<$ 50 kJ mol$^{-1}$ nm$^{-1}$. {\it emstep} = 0.001 is initial step size for position in  nm.  {\it nstcomm}= 1 is  the frequency for a center of mass motion removal, {\it ns\textunderscore type = grid } is to make a grid in the box and only check atoms in neighboring. {\it rlist}= 1.0 is fixed for cut-off distance for the short-range neighbor list in  nm. {\it coulombtype } = PME is used for the treatment of long-range electrostatic interactions, {\it vdwtype} = cut-off and {\it rcoulomb}=1.0 and {\it rvdw}=1.0 are for Coulomb and van der Waals cut-off. {\it nstxout}   = 200 represents the number of steps that elapse between writing coordinates. In this step, our system is not coupled with thermostat and barostat to maintain temperature and pressure because of fixed box size. {\it gen-vel  = no} represents the velocities are set to zero when there are no velocities in the input structure file. FIG.\ref{enm2} represents  our system  after energy minimization (em); after energy minimization molecules in box remain in more stable than before. \\

\noindent Energy minimization makes our system ready for the MD run. The physical properties of the system in equilibrium are not depended to initial conditions then equilibration run is performed for this. It helps to study dynamical variable which is always altered with different parameters like temperature, pressure, density, etc., in this step system should be coupled to thermostat and barostat. Which helps to bring to the state of thermodynamical equilibrium. Then we perform production run.  Thermal equilibrium and constant pressure are  obtained by coupling  of the system with a suitable thermostat and barostat called temperature and pressure coupling respectively used to rescale their respective parameters.\\

	\noindent	PME (Particle Mesh Ewald) type is used for long-range Coulomb interaction with Fourier-spacing  of 0.12 nm grid spacing for FFT with cut-off distance of 1.0 nm. {\it nslist = 10 } is the  frequency to update the neighbor list. {\it ns \textunderscore type = grid} is used to search  neighboring grid cells in non-bonded interaction within 1.0 nm of reference particle. {\it v-rescale} is used for temperature coupling i.e., modified Berendsen thermostat.  Different runs are carried  out at five different temperatures 293.20 K, 303.20 K, 313.20 K, 323.20 K and  333.20 K. The parameter {\it ref \textunderscore t} is the reference temperature; which, is set according to the different temperature at which simulation carried out. By fixing 0.01 ps time constant for temperature coupling.  In similar pattern pressure coupling is also fixed by Berendsen barostat  to  1 bar as a  reference pressure with  the time constant  0.8 ps; for water at 1 atm pressure at 300 K, isothermal compressibility is 4.6$\times$10$^{-5}$  bar$^{-1}$.  Velocities are generated  according to Maxwell distribution at different temperatures.  During equilibration run, all bonds are converted into constraints using constraint algorithm LINCS. \\

	\noindent	 When the system is equilibrated by  the {\it NPT} ensemble then we start to production run. After {\it NPT} ensemble density, temperature, and pressure of our system are fixed.  In this step, we apply {\it NVT} ensemble to study diffusion coefficient of Val. The pressure is not needed to couple: we don't use velocity generation because initial velocities are used as what generated in equilibration. LINCS algorithm is used in the production run. \\

		\section*{Results and Discussion}
		\noindent In this section, we discuss the structural  properties and diffusion coefficient  of the constituents of the systems. 
		\section*{Structure of the System}
		\noindent To analyze the structure of   our system we do study about radial distribution function (RDF). It gives the ideas about how the atoms or molecules are packed with respect to reference atom or molecule. Structure of the solvent can be interpreted by RDF. It describes  how density and probability vary with distance.
		\subsection*{Radial Distribution Function of Solvent }
		
		\noindent  Within  the preferred simulation box,  RDF describes the equilibrium structure of the water molecules.
		In this simulation, we have taken  the SPC/E water model.  Hydrogen atoms of water do not take part in LJ interaction with any other atoms. We use RDF  {\it g$_{OW-OW}$} to study the structure of water molecule in this simulation.\\
		\begin{figure}[h!]
			\centering
			\includegraphics[scale=0.26]{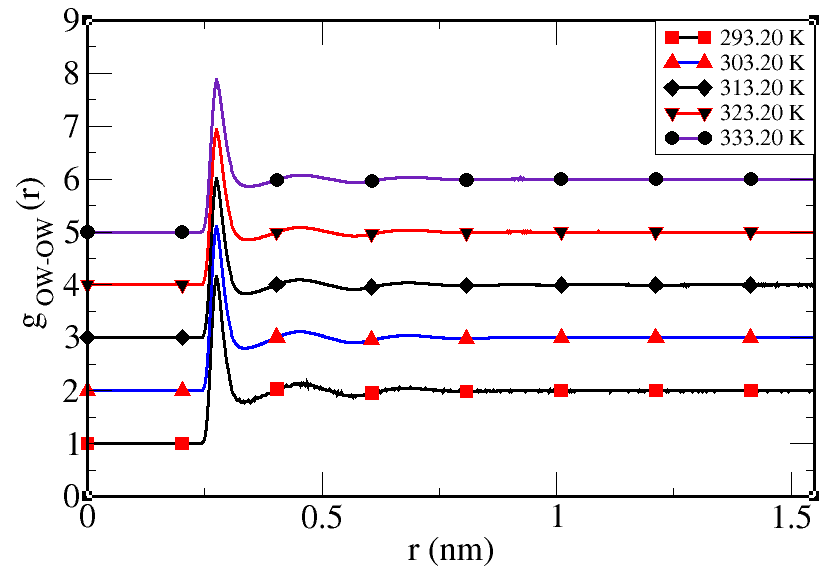}
			\caption{ Plot of  the RDF $g_{OW-OW} (r)$ i.e., Oxygen-Oxygen of water molecule  vs r(nm) at the different temperature. }
			\label{rdf1}
		\end{figure}

	\noindent	In FIG.\ref{rdf1}, we have drawn different RDF at corresponding temperatures. This gives the  probability of finding the oxygen atom of a water molecule at a distance r around the reference oxygen atom of the water molecule.
		In  FIG.\ref{rdf1}, reference atom is fixed at the origin. First peak point is seen which is very close to the reference  oxygen atom. The magnitude of the {\it g$_{OW-OW}$} is zero within atomic separation. Similarly, the second peak does contain  the second closest oxygen atoms to reference oxygen atom. Between first and second peak, there is the lowest region around $r\approx0.334$ nm, this notifies that, there is very low probability to  find oxygen atoms within this region. These peaks represent coordination shells. As in figure after certain distance i.e., beyond  the third peak,  the RDF is a straight line and it appears to be 1 means  there exists almost no pair correlation. As shown in table \ref{rdftable1} first, second and third peak points and  their corresponding values are measured. For example, at 293.20 K first peak point is at 0.2740 nm with corresponding value (height)  3.1514, second peak point at 0.4520 nm with corresponding value  1.1398 and so on. It means  that the closest  atoms can be found is at distance 0.2740 nm and it is  $3.15$  times more possibilities that two molecules would have existed at this point such that first coordination shell of  the oxygen atom of water around reference oxygen atom of water at this temperature lies at distance 0.2740 nm.

		\begin{table}[H]
			\centering
			\caption{Detail of RDF g$_{OW-OW}$ (r) at different temperatures.}
			\label{rdftable1}
			\scalebox{0.65}{
				\begin{tabular}{|l|l|l|l|l|l|l|l|}
					
					\hline
					
					Temp. &FPP(nm)&FPV&SPP(nm)&SPV&TPP(nm)&TPV&Co-ordination no.\\
					\hline
					293.20 K&0.2740&3.1514&0.4520&1.1398&0.6880&1.0566&421\\
					\hline
					
					303.20 K&0.2760&3.0987&0.4500&1.1176&0.6820&1.0430&406\\
					\hline
					313.20 K&0.2760&3.0073&0.4500&1.0983&0.6820&1.0395&399\\
					\hline
					323.20 K&0.2760&2.9360&0.4500&1.0855&0.6900&1.0361&395\\
					\hline
					333.20 K&0.2760&2.8806&0.4500&1.0778&0.6860&1.0360&393\\
					\hline

				\end{tabular}
			}
		\end{table}

	\noindent	From table \ref{rdftable1}, we see that when the temperature increases then the peak height decreases, and width of peak increases. This means  the solvent becomes less structural. Thermal agitation of atoms on the system also increases with increase in temperature.  Up to second decimal, all the  respective peak positions are  the same for all temperatures.\\
		
	\noindent	Co-ordination number  is the total no of atoms which are found around the reference atom. In RDF first minima suggests there are possible first co-ordination numbers and this can be calculated by FORTRAN code, then we get first coordination number. The integral equation can be discretized over radial distance as,
		\begin{equation}
		\centering
		\label{rdfdiscrete}
		\sum_{r_i=0}^{r_i=r_o} 4\pi r_i^2 \rho (r)  \mathnormal{\Delta} r.		\end{equation}
		In our case, step size of r ($\Delta r$) = $0.002$ nm.  First coordination numbers are calculated and expressed in table \ref{rdftable1}.  It suggests that when the temperature  increases with a corresponding decrease in density the coordination number decreases.\\

	\noindent	From {\it ffnonbonded.itp } file we see $\sigma$ for oxygen atom of water is 0.32 nm  such that van der Waals radius ($2^{\frac{1}{6}}\sigma$) is approximately 0.36 nm as shown in FIG.\ref{ljcl11}(up). The FPP is less than van der Waals radius-- suggests that there are other factors  apart from van der Waals interaction.\\

		{ \color{black} \begin{figure}[h!]
				\centering
				\begin{minipage}[b]{0.49\textwidth}
					\includegraphics[width=\textwidth]{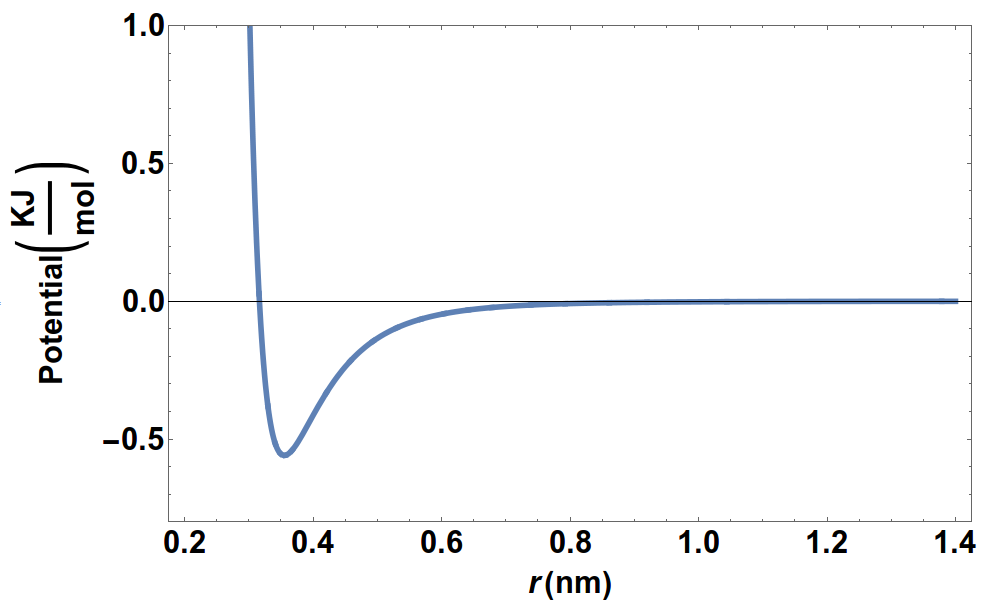}
					\
				\end{minipage}
				\hfil
				\begin{minipage}[b]{0.5\textwidth}
					\includegraphics[width=\textwidth]{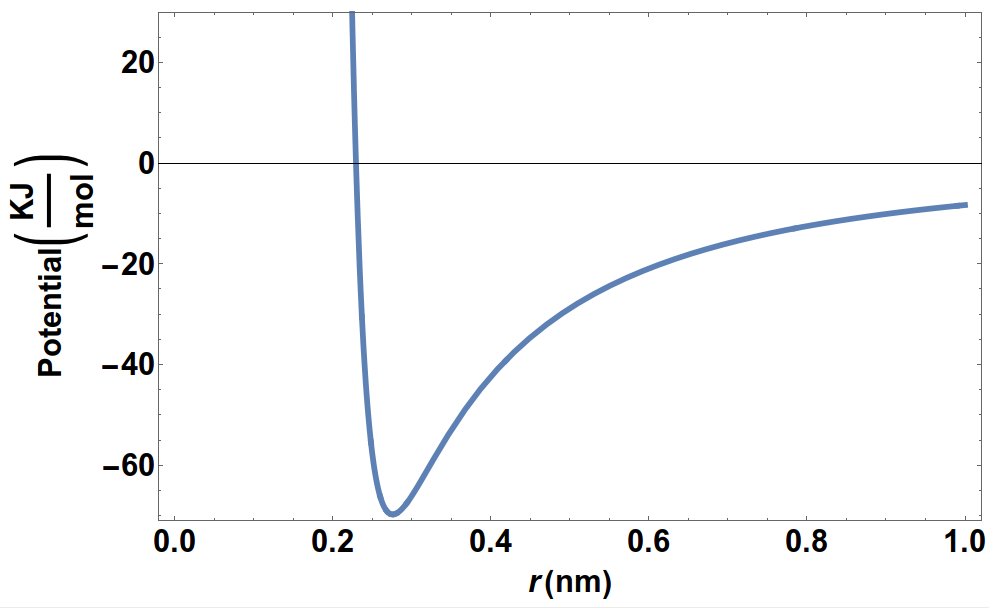}
					
				\end{minipage}
				\caption{LJ-(SR)-Potential(up) and  LJ-(SR)+Coulomb(SR)-Potential(below)  between OW-OW of water at 293.20 K.}
				\label{ljcl11}
			\end{figure} }

\noindent	The shift in the minimum potential energy position of water when Coulomb potential
			is added to LJ potential is illustrated by the FIG.\ref{ljcl11}(below). Our system incorporates
			other interactions like bonded and many-body effects, so this minimum position shifts again towards the origin. The minimum potential energy position should lie at the first peak position of the RDF.    
			\subsection*{Radial Distribution Function of OW-N6 }
\begin{figure}[h!]
				
				\centering
				\includegraphics[scale=0.26]{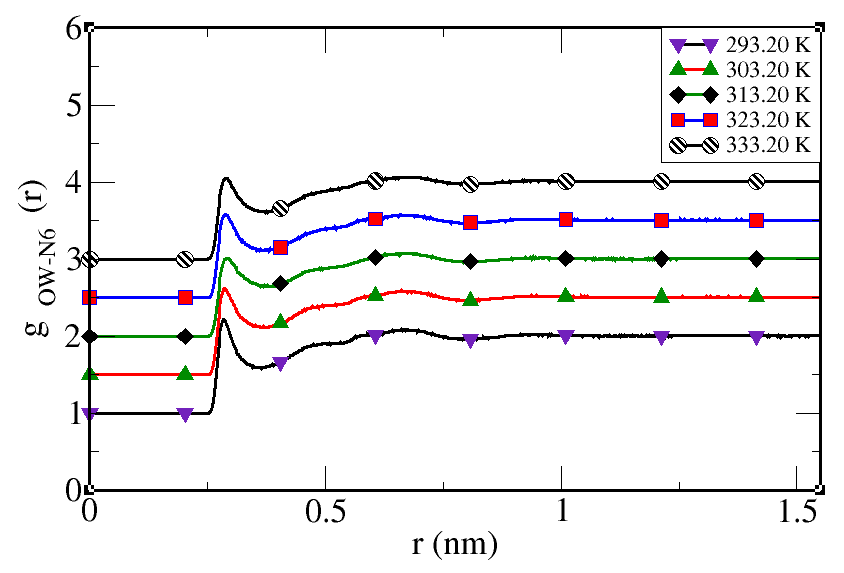}
				\caption{ Plot of  the RDF $g_{N6-OW} (r)$ i.e.,  Nitrogen atoms of Val and Oxygen of water molecules g(r) vs r(nm) at different temperature.  }
				\label{rdf2}
			\end{figure}

			\noindent In this system, N6  atom is the nitrogen atom of the amine group of Val and  OW  is oxygen atom of water: both have negative charge in nature. From the FIG.\ref{rdf2} we see that there exists first peak value 1.22 at position 0.28 nm at 293.20 K. It means that  there is two or more probability to find oxygen atoms of water around N6 of Val, and  radius of first coordination shell of  oxygen atom of water around the reference nitrogen atom of Val molecule is FPP of RDF. 
			We  can easily see the second peak but the third peak is quite difficult to observe. Both peaks have large width but less height  and appear to be 1 after certain distance r, it reflects there is no pair correlation.\\

		\noindent	Table \ref{rdftable2}  represents   the details of peak positions, it's value  and  coordination number of FIG.\ref{rdf2}, we see that when the temperature increases  the height of peaks decreases, width increases, and coordination number decreases.

			\begin{table}[H]
				\centering
				\caption{Detail of RDF g$_{N6-OW}$ (r) at different temperatures.}
				\label{rdftable2}
				\scalebox{0.67}{
					
					\begin{tabular}{|l|l|l|l|l|l|l|l|}
						
						\hline
						
						Temp. &FPP(nm)&FPV&SPP(nm)&SPV&TPP(nm)&TPV&Co-ordination no.\\
						\hline
						293.20 K&0.2840&1.2200&0.6720&1.0860&0.9280&1.0240&113\\
						\hline
						
						303.20 K&0.2860&1.1200&0.6620&1.0920&0.9480&1.0230&105\\
						\hline
						313.20 K&0.2920&1.0090&0.6800&1.0780&0.9560&1.0200&103\\
						\hline
						323.20 K&0.2880&1.0770&0.6600&1.0740&0.9550&1.0200&101\\
						\hline
						333.20 K&0.2880&1.0528&0.6840&1.0644&0.9380&1.0193&99\\
						\hline

					\end{tabular}
				}
			\end{table}

				\noindent The value of $\sigma$ for OW-N6 is 0.31 nm. The corresponding van der Waals radius is 0.35 nm.
			 Table \ref{rdftable2} shows that the FPP is less than van der Waals radius-- suggests that there are other factors  apart from van der Waals interaction like as Coulomb interactions, bonded interactions  and many-body effects.

				\subsection*{Radial Distribution Function of OW-H19 }
\begin{figure}[h!]
					\centering
					\includegraphics[scale=0.26]{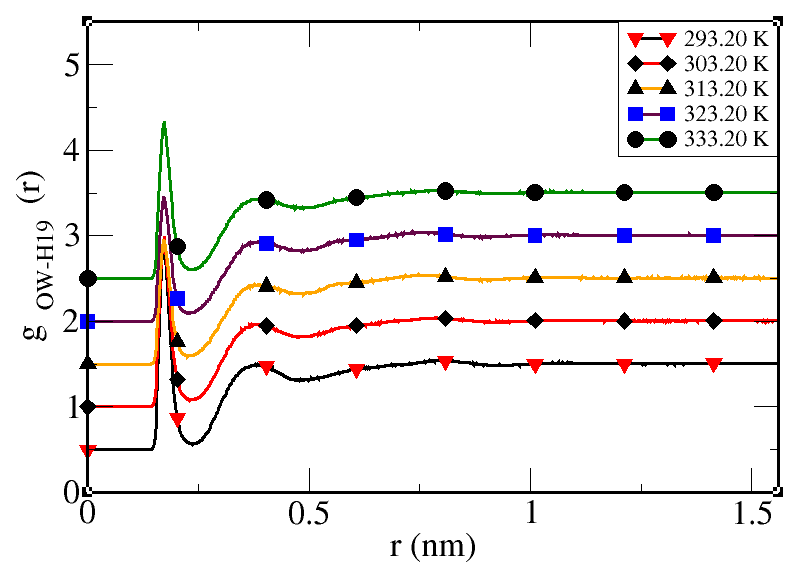}
					\caption{ Plot of  the RDF $g_{H19-OW} (r)$ i.e.,   Oxygen of water  and Hydrogen   from Carboxylic acid group g(r) vs r(nm) at different temperature.
					}
					\label{rdf4}
				\end{figure}

				\noindent  In this system, H19 is the hydrogen atom which is singled bonded with  the oxygen atom of the carboxylic group  and OW is oxygen of water molecule. FIG.\ref{rdf4} and table \ref{rdftable4} show that the different peak values and positions. At temperature 293.20 K, FPP is 0.17 nm with corresponding value  2.44, SPP is $\approx 0.38$  nm with corresponding value 0.99 and TPP is 0.80 nm with a corresponding value 1.04 after that no peaks appears so there is no pair correlation. The first peak value has the highest among other peaks so in this position, these atoms prefer to remain from each other. FPP is also can be considered as  radius of  the first coordination shell. We see that when the temperature increases peak value slightly alters and it's width increases this is supposed to due to thermal agitation and bonded or non-bonded interactions.\\

				\begin{table}[H]
					\centering
					\caption{Detail of RDF g$_{H19-OW}$ (r) at different temperatures.}
					\label{rdftable4}
					\scalebox{0.75}{
						\begin{tabular}{|l|l|l|l|l|l|l|}
							
							\hline
							
							Temp. &FPP(nm)&FPV&SPP(nm)&SPV&TPP(nm)&TPV\\				
							\hline
							293.20 K&0.1720&2.4350&0.3780&0.9960&0.8000&1.0430\\
							\hline
							
							303.20 K&0.1720&1.9750&0.3960&0.9650&0.7860&1.0410\\
							\hline
							313.20 K&0.1700&1.4440&0.3870&0.9300&0.7460&1.0420\\
							\hline
							323.20 K&0.1720&1.8950&0.3880&0.9480&0.7860&1.0370\\
							\hline
							333.20 K&0.1720&1.8296&0.3920&0.9279&0.7880&1.0321\\
							\hline
						\end{tabular}
					}
				\end{table}

			\noindent The value of $\sigma$ for OW-H19 is 0.00 nm. The corresponding van der Waals radius is 0.00 nm.
			Table \ref{rdftable4} shows that the FPP  lies at 0.17 nm-- suggests that  other factors  apart from van der Waals interaction like as Coulomb interactions, bonded interactions  and many-body effects are highly dominated.

					\subsection*{Radial Distribution Function of OW-C1 }
\begin{figure}[h!]
						\centering
						
						\includegraphics[scale=0.27]{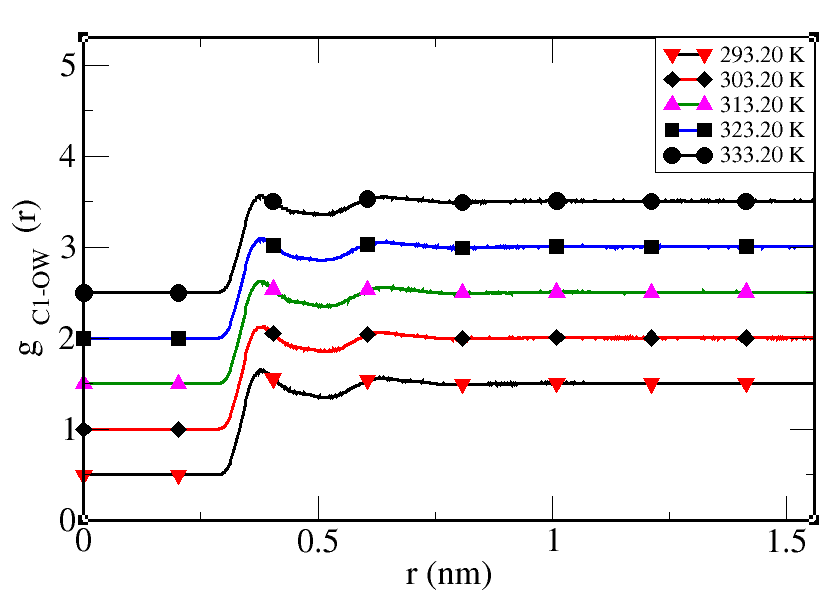}
						\caption{ Plot of  the RDF $g_{C1-OW} (r)$  i.e.,   Oxygen of water and Carbon from R-chain g(r) vs r(nm) at different temperature. 
						}
						\label{rdf5}
					\end{figure}

					\noindent In our system, C1 is  a carbon atom from R-chain of Val molecule. The RDF is taken between C1 and OW. Both have the negative charges but different in magnitude, they have repulsive Coulomb interaction between them.  Such that RDF of both are identical. From FIG.\ref{rdf5} and table \ref{rdftable5}  we can see those different properties as already discussed. At temperature 293.20 K, FPP is  $\approx0.38$ nm is a radius of first coordination shell and corresponding value 1.16 means that at this position there is 1.16 times chance to find oxygen atom of water around C1 atom of Val.  As in table \ref{rdftable5}, when the temperature increases the peak values decreases, width increases and coordination number decreases.  The third peak is difficult to observe but after that, we could not find any observable peak  which equals to 1 refers there is no pair correlation  beyond this region.\\

					\begin{table}[H]
						\centering
						
						\caption{Detail of RDF g$_{C1-OW}$ (r) at different temperatures.}
						\label{rdftable5}
						\scalebox{0.67}{
							\begin{tabular}{|l|l|l|l|l|l|l|l|}
								
								\hline
								
								Temp. &FPP(nm)&FPV&SPP(nm)&SPV&TPP(nm)&TPV& Co-ordination no.\\				
								\hline
								293.20 K&0.3780&1.1570&0.6400&1.0670&0.9960&1.0140&437\\
								\hline
								
								303.20 K&0.3850&1.1340&0.6360&1.0660&1.0080&1.0160&424\\
								\hline
								313.20 K&0.3780&1.1280&0.6320&1.0640&1.0640&1.0140&418\\
								\hline
								323.20 K&0.3760&1.0950&0.6460&1.0590&1.0650&1.0150&397\\
								\hline
								333.20 K&0.3800&1.0646&0.6380&1.0537&1.0020&1.0146&391\\
								\hline

							\end{tabular}
						}
					\end{table}
						
						\noindent The value of $\sigma$ for OW-C1 is 0.33 nm. The corresponding van der Waals radius is 0.37 nm. Table \ref{rdftable5} shows that the 
						FPP is less than van der Waals radius-- suggests that there are other factors  apart from van der Waals interaction like as Coulomb interactions, bonded interactions  and many-body effects. 
						
						\section*{Diffusion Coefficients}
						
						\subsection*{Self Diffusion Coefficient of Val}
\begin{figure}[H]
							\centering
							\includegraphics[scale=0.28]{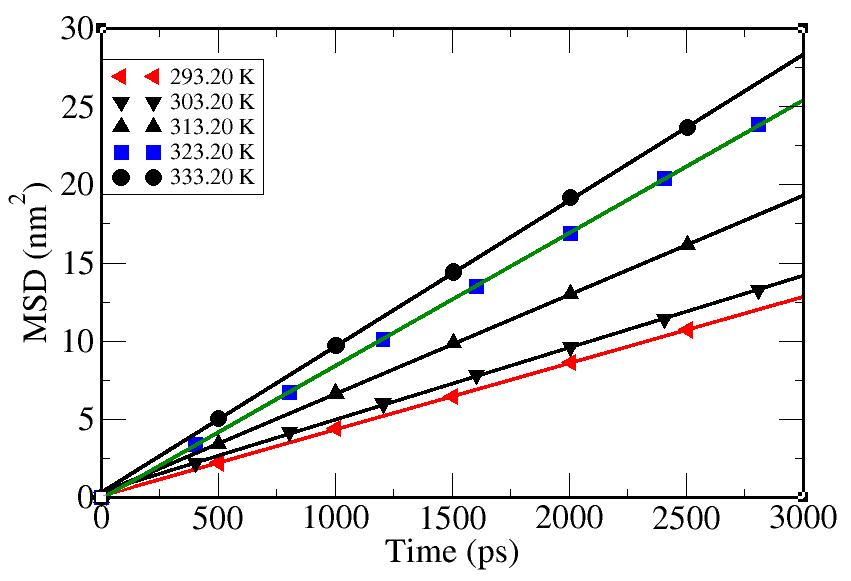}
							\caption{MSD plots for Val obtained from simulation and their linear fit at different temperatures.}
							\label{msdvalall}
						\end{figure}

						\noindent	We have estimated self-diffusion coefficient $D_{Val}^{Self}$ of Val at 293.20 K,   303.20 K, 313.20 K, 323.20 K and 333.20~K.    FIG.\ref{msdvalall} shows the MSD vs time of Val  for 3 ns  and their repective linear fit at different temperatures,  It is found that  when temperature increases the slope of MSD vs time also increases, such that the self-diffusion coefficient of Val also increases with increase in temperature. It is seen that self-diffusion coefficient of  Val at 333.20 K is   120 \% greater than  at 293.20 K. \\

						\subsection*{Self Diffusion coefficient of Water}

\begin{figure}[H]
							\centering
							\includegraphics[scale=0.33]{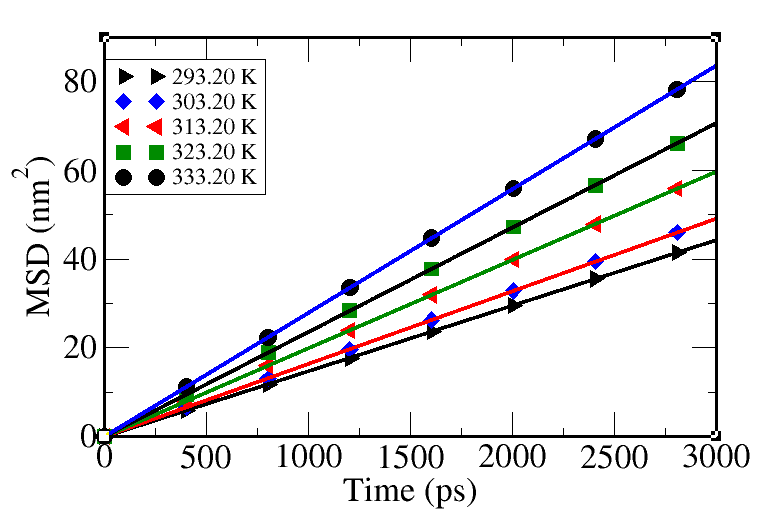}
							\caption{MSD plot of water at different temperature.}
							\label{msdwaterall}
						\end{figure}

						 We have estimated self-diffusion coefficient$D_{water}^{Self}$  of water at   293.20 K, 303.20 K, 313.20 K, 323.20 K and 333.20 K.    FIG.\ref{msdwaterall} shows the MSD vs time of water  for 3 ns  and their respective linear fit at different temperatures,  It is found that  when temperature increases the slope of MSD vs time also increases, such that self-diffusion coefficient of water also increases with increase in temperature. It is seen that self-diffusion coefficient of  water at 333.20 K is   111.50 \% greater than at 293.20 K. \\
						
						\subsection*{Binary Diffusion Coefficient of Val-water System}
						
						\noindent During this simulation, we have used 3 Val molecules and 1047 water molecules. Thus, the mole fraction of Val and water are  0.003 and 0.997 respectively. In this case, Val molecules are very much smaller than water molecules such that the mole fraction of water dominates at all and it's coefficient in equation \ref{darken} i.e.,  $D_{Val}^{self }$ is much  closer to the binary-diffusion coefficient. \\

				\begin{table}[H]
							\caption{Binary - diffusion coefficient of Val in water at different temperatures.}\label{msdmixdata}
							\resizebox{0.480\textwidth}{!}{
								\begin{tabular}{|c|c|c|c|}
									\hline
									
									Temp. &Binary-Diffusion & Experimental Value&Error \\
									& \small{($D_{simul.}^{binary}$ in $10^{-10}$) m$^2$/sec}&\small{($D_{exp.}^{binary}$ in $10^{-10}$) m$^2$/sec}&\%\\
									& & \cite{15}&\\
									\hline
									293.20 K&7.12 $\pm $ 0.30& 6.76 $\pm $ 0.11&5.32 \\
									\hline
									
									303.20 K&7.74 $\pm $ 0.20&8.91 $\pm $ 0.06&13.13\\
									\hline
									313.20 K&10.65 $\pm $ 0.98&11.24 $\pm $ 0.10&5.25\\
									\hline
									323.20 K&14.23  $\pm $ 0.89&14.02 $\pm $ 0.15&1.50\\
									\hline
									333.20 K&  15.65  $\pm $ 0.25&17.03 $\pm $ 0.16&8.10\\
									\hline

								\end{tabular}
							}
						\end{table}

							\noindent	Table \ref{msdmixdata} represents all the simulated values of binary-diffusion coefficient of Val-water mixtures after calculating from Darken's relation at different temperatures. And we compared available experimental result from \cite{15}. Our work in simulation is quite close with experimental result spite of being infinite dilution within the error  ranging from  1.50 \%  to 13.13 \%.  \\

						 \begin{figure}[h!]
						 	\centering
						 	\includegraphics[scale=0.283]{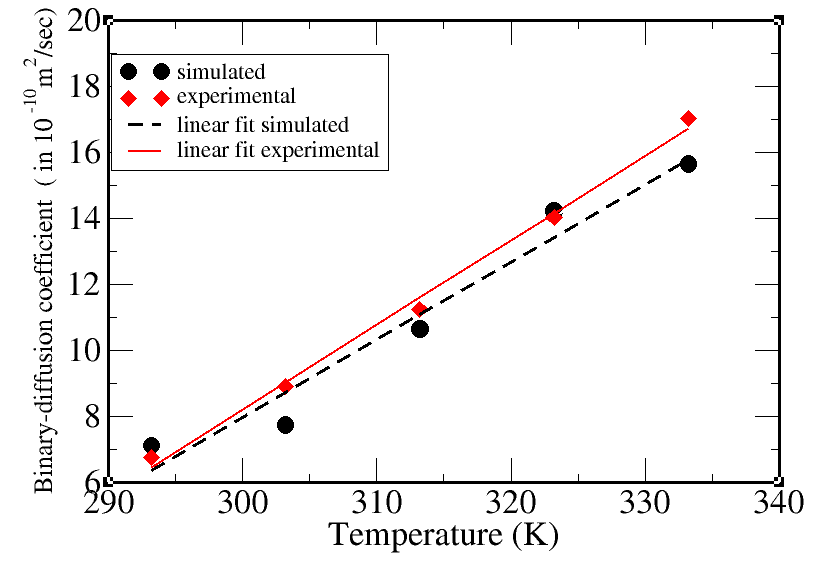}
						 	\caption{ Variation of  the binary-diffusion coefficient with temperature obtained from simulation and experimental literature\cite{15}.}
						 	\label{binarydiffusion}
						 \end{figure}
						
						\noindent In  FIG.\ref{binarydiffusion}, we plotted our simulated result with experimental result found in literature, the black circle points represent our simulated result and diamond square points represent  the experimental result. We see that both results are quite close together  within some error. From this figure, we can conclude that the binary-diffusion of Val and water increases with increase in temperature.
						
						\section*{Temperature Dependency of Diffusion}
			\noindent	All the  graphs and tables of  self-diffusion and binary-diffusion show that the diffusion process proportional to the temperature, diffusion increase with an increase in temperature and vice-versa. This relation between temperature and diffusion follow Arrhenius formula\cite{1111}.
					\begin{equation}
					\label{arrhenius}
					ln(D)= ln(D_o) -\frac{E_\alpha}{N_A k_B T}
					\end{equation}
					The pre-exponential factor  $D_o$ is  the frequency factor, $E_\alpha$ is activation energy for diffusion, T is absolute temperature, $N_A=6.02 \times 10 ^{23}$ per mol is Avogadro number, and $k_B= 1.38\times10^{-23}$ JK$^{-1}$  is  Boltzmann constant.
					When we plot  a graph between  natural log (ln) of diffusion coefficient vs reciprocal of  absolute temperature is called  the Arrhenius diagram. The  $-N_A k_B$ times slope of this  graph  corresponds activation energy of diffusion process which can be expressed as,
					\begin{equation}
					\label{activation}
					E_\alpha = - N_A k_B  \frac{\partial\ln (D)}{\partial (1/T)}
					\end{equation}

					\noindent the pre-exponential factor can be  obtained when we extrapolate the intercept of this equation to the  1/T~$\rightarrow 0$.
					   \begin{figure}[h!]
					   	\centering
					   	\includegraphics[scale=0.28]{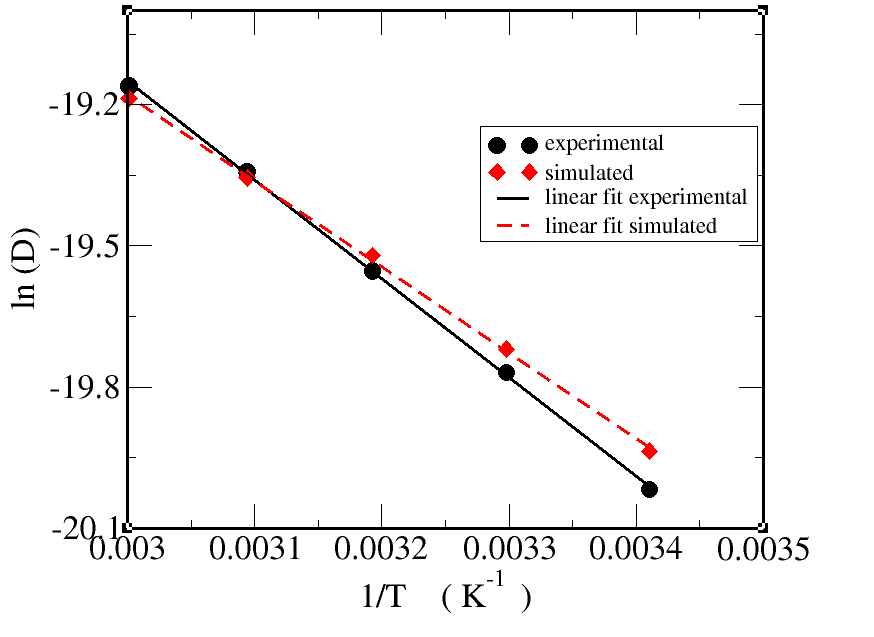}
					   	\caption{ Arrhenius diagram for the self-diffusion coefficient of water both simulated and experimental. }
					   	\label{selfwaterboth}
					   \end{figure} 
					  
					\noindent The  FIG.\ref{selfwaterboth} is Arrhenius diagram of self-diffusion of water for both simulation and experimental values,  corresponding activation energies are 15.14~kJ~mol$^{-1}$ and  17.42~kJ~mol$^{-1}$ respectively. We see that activation energy is in good agreement with the experimental activation energy with error 13.07 \%. This is represented in table \ref{act-table}.\\
					    
				\pagebreak	  
					  
						  \begin{figure}[h!]
						  	\centering
						  	\includegraphics[scale=0.28]{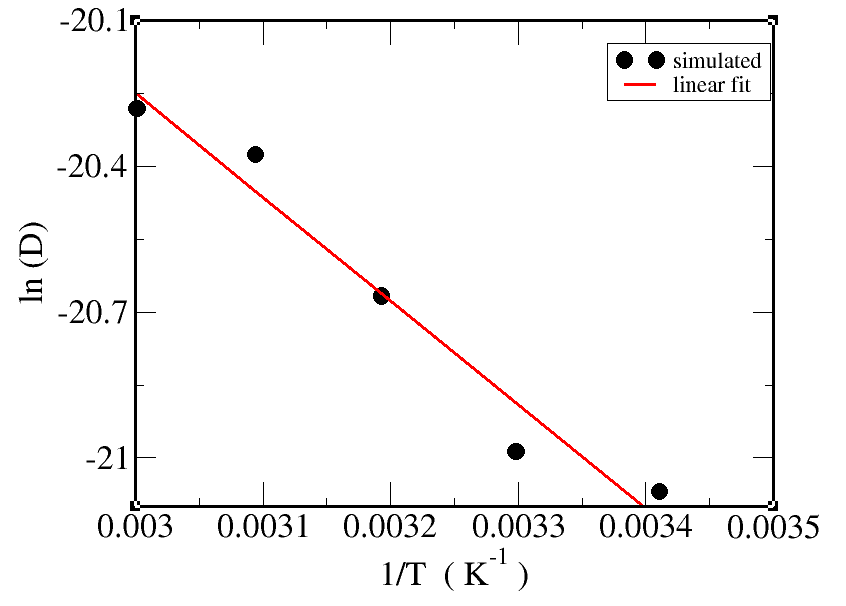}
						  	\caption{ Arrhenius diagram for the self-diffusion coefficient of Val. }
						  	\label{selfvaline}
						  \end{figure} 
						  
					\noindent	The  FIG.\ref{selfvaline} is Arrhenius diagram of self-diffusion of Val, so activation energy  from this diagram for   self-diffusion of Val is calculated, which is  ~17.74~kJ~mol$^{-1}$.\\
						  
					 \begin{figure}[h!]
					 
					 	\includegraphics[scale=0.26]{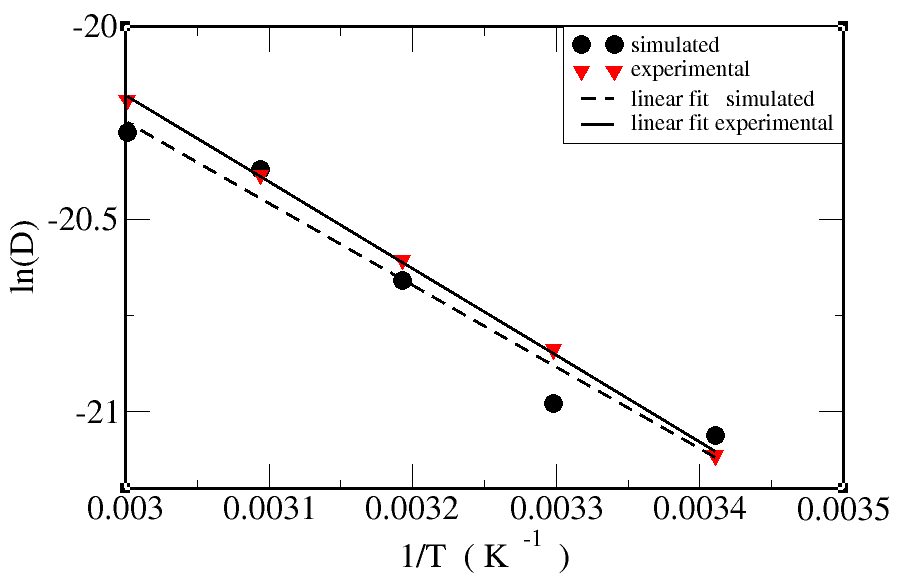}
					 	\caption{ Arrhenius diagram for the binary-diffusion coefficient of Val-water both simulated and experimental. }
					 	\label{binaryboth}
					 \end{figure} 
				
					 	\noindent		The FIG.\ref{binaryboth} is Arrhenius diagram of binary-diffusion of Val-water for both simulation and experimental values, so activation energies  from this diagram  are calculated  for both case, which are equal to 17.72 ~kJ mol$^{-1}$ and 18.72 ~kJ mol$^{-1}$ respectively. We see that activation energy is in good agreement with the experimental activation energy with error 5.34 \%.
					 Table \ref{act-table} represents activation energies with corresponding species.\\	\begin{table}[H]
							\centering
							\caption{Pre-exponential factor and activation energy of diffusion.}
							\label{act-table}
							\label{table:activationenergy}
							\resizebox {0.48\textwidth }{!}{%
							\begin{tabular}{|c|c|c|c|}
								\hline
								System  & \multicolumn{2}{|c|}  {Activation energy (E$_a$) in kJ mol$^{-1}$} & Error Percentage\\
								\hline
								& Simulated & Experimental & \\
								\hline
								Water-(self) &15.14 &  17.42 \cite{selfwater}&13.07\\
								\hline
								Val-simulated (self)&17.74&---&---\\
								\hline
								Binary mixture &17.72 &  18.72  \cite{15}&5.34\\
								\hline
							\end{tabular}}
						\end{table}

						\section*{Conclusions and Concluding~Remarks}
						\noindent We have performed this work under classical molecular dynamics of   binary mixture of 1047 water molecules of species SPC/E  model  and 3 Val molecules  at different temperatures 293.20 K, 303.20 K, 313.20 K, 323.20 K, and  333.20 K. The  molar concentration was fixed to   0.003   for all the temperatures i.e., almost 160 mol m$^{-3}$, which was  considered as infinite dilution.\\
						
						\noindent		We have used GROMACS 4.6.5 package for this simulation in UBUNTU 14.04 environment and we analyzed all the obtained data  by XMGRACE and VMD.
						Bonded interactions like an angle, dihedral were taken into account whereas bond vibration was constrained.  All the short-range non-bonded interactions  were taken into account and long-range Coulomb interaction was treated  by PME and LJ interaction was restricted  by cutoff within 1.0 nm. 
						To get the initial minimum energy configuration free from overlaps,  energy minimization was done by using  the steepest-descent algorithm. This step was found to converge well within the  specified tolerance force constant i.e., 50 kJ mol$^{-1}$ nm$^{-1}$.  Equilibration   was done for 100 ns using NPT ensemble   by coupling with velocity rescaling thermostat and Berendsen barostat, which was followed by production run,  where production run  was simulated for 100 ns    using  NVT ensemble.  \\
						
					\noindent			Different energy profiles  along with density profiles of the system were studied at different temperatures which were done to know the equilibrium nature of the system. The RDF  between different atoms like OW-OW, H19-OW, C1/C3-OW, N6-OW  were studied  and discussed to know the equilibrium structural properties of the system,  Different peaks in RDF allowed to  know  how atoms were structured  in the system,  Those peaks helped to analyze the liquid system is more ordered than gases but random to  crystalline solid.\\
						
						\noindent		The equilibrium dynamical property of the system called the transfer of mass--diffusion  coefficient was studied.  Concentration gradient was  the main factor for diffusion. The self-diffusion coefficient of Val and water at different temperatures were calculated independently by using  the MSD method.  Then binary diffusion coefficient of Val-water was calculated by using Darken's relation.
						Both self-diffusion of water and binary diffusion of binary solution were in good agreement with experimental data within error ranging from 1.41 \% to 8.54 \% and 1.50 \% to  ~13.13~\% respectively.  The temperature dependence of the diffusion coefficients  manifested   to follow Arrhenius behavior. 
						Activation energies for different species were calculated from Arrhenius diagram. \\

						\noindent		This study  helps to make the conclusion that  molecular dynamics simulation is one of the best methods to study equilibrium structure and transport properties of biomolecules.  Dynamic properties like as diffusion of  the binary mixture--which is quite reliable agreement with the experimental result \cite{15}. It is  economically and  computationally  less expensive.  There are many cases where experimental research  is difficult to carry out because of  economically expensive. This is quite feasible in our country Nepal where experimental researches are limited due to different cost.
						For the future study of biomolecules, we can use this calculated diffusion coefficient and structural properties. \\

						\noindent		In future, we can extend our work in many ways  adding transport properties  like viscosity and thermal conductivity to our work for different concentrations. Zwitterion of Val at different temperatures  and concentrations can be studied.  Even allowing long ranged van der Waals interaction can be another research in the extension of this work, Hard water can be used instead of water.
						We can study interaction of Val molecules with advanced materials like graphene, carbon nanotube,
						Nano-bio interactions between carbon nanomaterials and blood plasma proteins is also another field of research.\\
							\section*{Acknowledgements}
						\noindent D. Pandey acknowledges the financial support from University Grants Commission (UGC), Nepal. NPA also
						acknowledges UGC Nepal award No. CRG73/74-S\&T~01. We also acknowledge the computational facility by ICQ-13.


\begin{thebibliography}{10}
						
					
						
						
				
					
						
						%\bibitem{aminoacidfunction2}
				%	http://www.aminoacidstudies.com,
				%		\newblock What are amino acid (2018).
						
						
						
							\bibitem{dpk10}
							A. Panagopoulou,
							\newblock  Molecular Dynamics and phase transitions in protein-water
							system,
							\newblock PhD thesis, National Technical University of Athens faculty of  applied mathematics and physics (2013).
							
							
						\bibitem{dp}
						http://www.aminoacidsguide.com (2018).
						
						
					
						
						\bibitem{lvaline}
				https://pubchem.ncbi.nlm.nih.gov/compound/6287	(2018).	
					
							
							%\bibitem{dpk55}
							%https://www.vitaminstuff.com/amino-acid valine.htm,
							%\newblock Amino acid,valine  (2018).
							
							\bibitem{bb}
							F. Zhao,  and  ~et ~al.,
							\newblock Genetic analysis of pathway regulation for enhancing branched-chain amino acid biosynthesis in plants,
							\newblock {\em the plant journal}, (1)  (2010).
							
							\bibitem{dpk123}
							https://toxnet.nlm.nih.gov,
							\newblock L-Valine (2018).
							
								\bibitem{vro}
								R.D.~Jenert~Renne, MS,
								\newblock The amino acid which helps for weight loss, L-Valine,
								\newblock {\em livestrong.com}, {\bf 1}(2) (2005).
							
							\bibitem{dpk13}
							M. T.  Farran, and  O. P. Thomas, Valine Deficiency, the effect of feeding a Valine--Deficient diet during the starter period on performance and leg abnormality of male broiler Chicks1 {\em Poultry Science},  http://dx.doi.org/10.3382/ps.0711885 (1992).
							
							
							
						
							
								\bibitem{x}
								R.B.~Russell and M.J.~Betts,
								\newblock  Amino acid protein and consequences of subsitutions in
								Bioinformatics for Geneticists, publisher={M.R. Barnes, I.C. Gray eds,  wiley}, http://www.russelllab.org/aas/Val.html  (2003).
						
							
							\bibitem {29}
							A. D. Jong and A.C. Borstlap,
							\newblock  Transport of amino acids (lvaline,llysine, Iglutamic acid) and sucrose into plasma membrane vesicles isolated from cotyledons of developind pea seeds,
							\newblock {\em Journal of Experimental Botany}, {\bf 51}(351) (2000).
							
							\bibitem{1}
							\newblock D. Allemand,  and ~et ~at.,
							\newblock Characterization of valine transport in sea urchin eggs,
							\newblock {\em Biochimica et Biophysica Acta (BBA)-Biotechnology Progress}, {\bf 772}(3)  (1984).
							
							
							\bibitem{4}
							\newblock J. Kotthaus,  and ~et ~at.,
							\newblock Synthesis and biological evaluation of l-valine-amidoximeesters as double progrugs of amidines,
							\newblock {\em Bioorganica \& medicinal chemistry}, {\bf 19}(6) (2011).
							
							\bibitem{24} 
							\newblock  P.L. Smith, and ~et ~at., 
							\newblock Transport of l-valine-acyclovir via the oligopeptide transporter in the human intestinal cell line, caco-2,
							\newblock {\em Journal of Pharmacology and Experimental Therapeutics}, {\bf 286}(3) (1998).
							
							\bibitem{5}
							\newblock M. Ning, and ~et ~at.,
							\newblock Variation in intestinal transport of l-valine in relation to age, 
							\newblock {\em Experimental Biology and Medicine}, {\bf 129}(3) (1968).
							
							
							
							
							\bibitem {12}
							\newblock  Y. Ma, and ~et ~at., 
							\newblock Studies on the diffusion coefficients of amino acids in aqueous solution,
							\newblock {\em Journal of Chemical and Engineering Data}, {\bf 50}(4)  (2005).
							
							
							
							
							\bibitem{15}
							T. Umecky, and  ~et ~al.,
							\newblock Infinite dilution binary diffusion coefficients of several
							$\alpha$-amino acids in water over a temperature range from (293.2 to 333.2)
							K with the taylor dispersion technique,
							\newblock {\em Journal of Chemical \& Engineering Data}, {\bf 51}(5)
							(2006).
							
						
							
							\bibitem{crank}
							J. Crank,
							\newblock  The mathematics of diffusion,
							\newblock {\em Oxford university press} (1979).
							
							
							
						
							
							\bibitem{abc}
							P.~Heitjans and J.~Krager,
							\newblock {\em Diffusion in Condensed Matter; Methods, Materials, Models},
							\newblock Number~1 Springer (2009).
							
							
							\bibitem{sunil}
							S.~K. Thapa,
							\newblock Molecular dynamics study of diffusion of oxygen in water at different
							temperatures,
							\newblock Master's thesis, Central Department of Physics, TU Kathmandu Nepal  (2010).
							
							\bibitem{aaa}
							
							\newblock { L.S. Darken,``Diffusion, Mobility and Their Interrelation
								through Free Energy in Binary Metallic Systems'', Trans. AIME, vol. 175, 1 (1948)}.
							
							%\bibitem{rdf1}
						%	J.~L. Yarnell,  and  ~et ~al.,
						%	\newblock Structure factor and radial distribution function for liquid argon at  85 K,
						%	\newblock {\em Phys. Rev. A}, {\bf 7}  (1973).
							
							
								\bibitem{smith} B.~Smit,  and ~et ~al.,
								\newblock {\em Understanding Molecular simulation from algoritm to
									apalicaltions}, volume~1,
								\newblock Academic Press, San Diego San Francisco New York Boston London Sydney
								Tokyo (1996).
						
						%	\bibitem{rdf2}
						%	D.~{Chandler} and J.~K. {Percus},
						%	\newblock {Introduction to Modern Statistical Mechanics},
						%	\newblock {\em Physics Today}, 41, 114 (1988).
							
							
							
						%	\bibitem{rdfwjb}
						%	W.J.~Briels,
						%	\newblock Theory of polymer dynamics,
						%	\newblock {\em Lecture Notes; Uppsala University; Uppsala, Sweden} (1994).
							
						
							
							
						
							
						
							
						%	\bibitem{kittel}
						%	C. Kittel,
						%	\newblock  Introduction to solid state physics, volume~8,
						%	\newblock Wiley New York (1996).
							
							\bibitem{allen}
							M.P.~Allen and D.J.~Tildesley.
							\newblock Computer simulation of liquids, volume 18 of oxford science
							publications,
							\newblock {\em Oxford University Press}, 45, 121 (1989).
							
						%	\bibitem{molecular}
						%	J.~Li,
						%	\newblock Basic molecular dynamics,
						%	\newblock In handbook of Materials %Modeling,  Springer
						%	(2005).
							
							\bibitem{bhakta}
							B.~R. Niraula,
							\newblock Molecular dynamics study of diffusion of propane gas in water at
							different temperatures,
							\newblock Master's thesis, Central Department of Physics, TU Kathmandu Nepal,
							Nepal (2017).
							
							
							
							
							\bibitem{usermanual}
							D.~V. Lindahl, and et~al.,
							\newblock Gromacs user manual version 4.6,
							\newblock {\em Search PubMed} (2013).
						%	\bibitem{reftemp}
					%\newblock 		%http://antoine.frostburg.edu/chem/senese/javascript/w%ater density  (2018).
							
							\bibitem{selfwater}
							M. Holz,  and ~et ~al.,
							\newblock ~Temperature-dependent~self-diffusion~coefficients~of ~water~and~six~selected~molecular~liquids~for~calibration~in~accurate~1H~NMR~PFG~measurements  ~(2000).
							
							\end{thebibliography}
					\end{document}